\documentclass[prl,floatfix,showpacs,amsmath,twocolumn]{revtex4}
\usepackage{amssymb}
\usepackage{graphicx}

\begin{document}

\newcommand{\be}{\begin{eqnarray}}
\newcommand{\ee}{\end{eqnarray}}
\newcommand{\bea}{\begin{eqnarray}}
\newcommand{\eea}{\end{eqnarray}}
\newcommand{\bma}{\begin{subequations}}
\newcommand{\ema}{\end{subequations}}
\def\RR{\mathbb{R}}
\def\E{\mathbf e}
\def\D{\boldsymbol \delta}

\title{Discrete entanglement distribution with squeezed light}

\author{B. Kraus and J. I. Cirac}
\affiliation{Max-Planck-Institut f\"ur Quantenoptik,
Hans-Kopfermann-Str. 1, Garching, D-85748, Germany.}

\date{\today}

\begin{abstract}
We show how one can entangle distant atoms by using squeezed
light. Entanglement is obtained in steady state, and can be
increased by manipulating the atoms locally. We study the effects
of imperfections, and show how to scale up the scheme to build a
quantum network.
\end{abstract}

\maketitle

Distributing entanglement among different nodes in a quantum
network is one of the most challenging and rewarding tasks in
quantum information. This may allow to extend quantum cryptography
over long distances by using quantum repeaters \cite{repeaters}.
Furthermore, it may lead to some practical applications in the
context of secret sharing \cite{sharing} or distributed quantum
computation \cite{distributed}. From the more fundamental point of
view, it may allow to perform loophole free tests of Bell
inequalities \cite{Bell}.

In a quantum network photons are used to entangle atoms located at
different nodes which store the quantum information. Local
manipulation of the atoms using lasers allows then to process this
information. In principle, one can construct quantum networks
using discrete \cite{Cirac} (qubit) or continuous variable
entanglement \cite{PK} (the one contained, for example, in
two--mode squeezed states \cite{GardinerZoller}). However, the
fact that Gaussian states cannot be distilled using Gaussian
operations \cite{Gaussno} may strongly limit the applications of
continuous variable entanglement in quantum networks and
repeaters.

\begin{figure}[t]
  \centering
  \includegraphics[width=\linewidth]{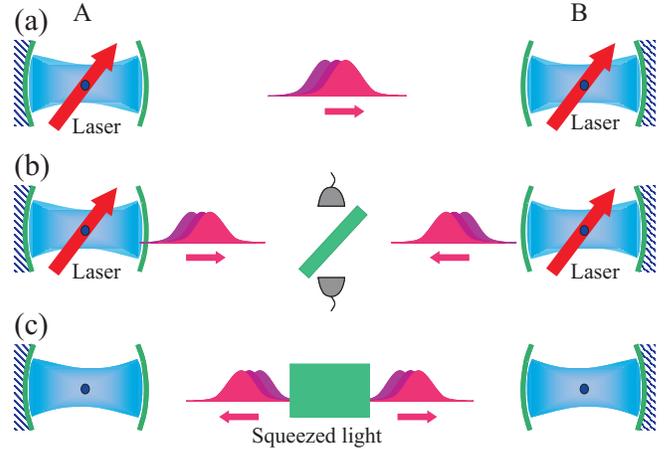}
  \caption{\label{fig1}
    Schemes for entanglement creation over long distances:
    (a) Entanglement is created by the emission and subsequent
    absorption of a photon;
    (b) A detection of a photon projects the atoms in an entangled
    state; (c) Both cavities are driven by a common source of
    two--mode squeezed light. In steady state, the atoms become
    entangled.}
\end{figure}

There have been several proposals to obtain discrete entanglement
of distant atoms using high-Q cavities. There are basically two
kind of schemes \cite{Cirac,Parkins1,Cabrillo,Huelga}: (1) [Fig.\
\ref{fig1}(a)] An atom A, driven by a laser, emits a photon into
the cavity mode. The photon, after travelling through a fiber,
enters the second cavity where it is absorbed by atom B, which is
also driven by a laser \cite{Cirac,Parkins1}. (2) [Fig.\
\ref{fig1}(b)] Both atoms are simultaneously driven by a laser in
such a way that if a photon is detected at half way between the
cavities, the atoms get projected into an entangled state
\cite{Cabrillo,Duan}. Most of these schemes operate in a
transitory regime; i.e., the entanglement is achieved at a
specific time and the lasers have to be switched on and off
appropriately. Moreover, dissipation may introduce imperfections
in the desired entangled state. In this work we propose and
analyze a scheme to distribute discrete entanglement which works
in steady state. As opposed to these other schemes, dissipation is
a necessary ingredient of our scheme which, as we will show, gives
it a very robust character. Our scheme transforms continuous
variable entanglement into discrete (qubit) entanglement and thus
exhibits how this last kind of entanglement may still be very
useful in the context of quantum networks. We show how a small
amount of this kind of entanglement can be used to create
maximally entangled qubit states. We also show how this scheme can
be scaled-up by using atoms with several internal levels.

The basic idea is schematically represented in Fig.\
\ref{fig1}(c). Both cavities are driven simultaneously by squeezed
light. The schemes ensures that part of the entanglement contained
in the light is transferred to the atoms. The use of squeezed
light to drive a single atom was first proposed by Gardiner
\cite{Gardiner}, who studied several phenomena on the atomic
steady state. Kimble and col. \cite{Kimble}, in a remarkable
experiment, were able to couple squeezed light in a cavity
containing atoms, and confirmed some of the physical phenomena
theoretically predicted. Recent experiments in which atoms have
been stored in high-Q cavities for relatively long times
\cite{Kimble3} pave the way for the implementation of several
quantum information protocols and, in particular, the one analyzed
in the present work.

Let us consider two two--level atoms, A and B, confined in two
identical cavities, which are separated by a certain distance. The
cavities are driven by an external source of two--mode squeezed
light [see Fig. \ref{fig1}(c)]. Assuming that the bandwidth of the
squeezed light is larger than the cavity damping rate $\kappa$,
the evolution of the atoms--plus--cavity modes density operator,
$\rho$, can be described using standard methods
\cite{GardinerZoller} by the following master equation
 \be
 \label{ME}
 \frac{d\rho}{dt}= -i[H_a + H_b,\rho]+ ({\cal L}_{\rm cav} +
 {\cal L}_{\rm at}^{a} + {\cal L}_{\rm at}^{b})\rho.
 \ee
Here $H_a=g_a (a\sigma_a^+ + a^\dagger \sigma_a^-)$ describes the
resonant interaction of atom A with the corresponding cavity mode,
where $a$ is the mode annihilation operator and
$\sigma_a^+=(\sigma_a^-)^\dagger=|e\rangle_a\langle g|$, with
$|g\rangle$ and $|e\rangle$ denoting the ground and excited atomic
states \cite{note}. Spontaneous emission
$|e\rangle_a\to|g\rangle_a$ is described by the usual Liouvillian
${\cal L}_{\rm at}^{a}$ \cite{GardinerZoller}, which is
proportional to the spontaneous emission rate $\Gamma$. The terms
$H_b$ and ${\cal L}_{\rm at}^{b}$ are analogously given. Finally,
the interaction between the cavity modes and the squeezed light is
given by
 \bea
 {\cal L}_{\rm cav} \rho &=& \kappa (N+1) \sum_{\alpha=a,b} (
 \alpha\rho\alpha^\dagger- \alpha^\dagger\alpha\rho) \nonumber\\
 &&+ \kappa N \sum_{\alpha=a,b} (
 \alpha^\dagger\rho\alpha- \alpha\alpha^\dagger\rho) \nonumber\\
 &&+ \kappa M (2 a\rho b + 2 b\rho a -2 ba\rho -2 ab\rho) + h.c.,
 \eea
where $h.c.$ denotes hermitian conjugate. Here, $N$ and $M$
characterized the two--mode squeezed vacuum and fulfil $M\le
N(N+1)^{1/2}$. In the following we will concentrate in the case
$g_a=g_b:=g$ since the formulas are considerably simplified. The
effects for the case $g_a\ne g_b$ will be analyzed at the end.

\begin{figure}[t]
  \centering
  \resizebox{\linewidth}{!}{
  \includegraphics{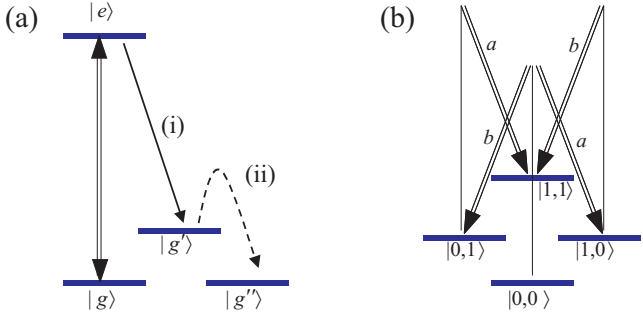}}
  \caption{\label{fig2}
    Atomic level schemes. Double lines indicates coupling to cavity modes and
    single lines to lasers: (a) Levels $|g\rangle$ and $|e\rangle$
    are used for entanglement creation; level $|g'\rangle$ is used to store the
    qubit once the entanglement has been created; level $|g''\rangle$ is used for
    entanglement concentration. (b) The four internal ground levels are coupled to
    two cavity modes ($a$ and $b$) by Raman transitions as indicated.}
\end{figure}

Let us first consider the ideal case in which $\Gamma=0$ and
perfect squeezing
 \be
 \label{perfectsq}
 M= [N(N+1)]^{1/2}.
 \ee
We can define new annihilation operators as $\tilde a=(N+1)^{1/2}
a + N^{1/2} b^\dagger$ and $\tilde b=(N+1)^{1/2} b + N^{1/2}
a^\dagger$, so that Eq.\ (\ref{ME}) can be rewritten as
 \be
 \label{ME2}
 \frac{d\rho}{dt}= -i[\tilde H_a + \tilde H_b,\rho]+ \tilde{\cal L}_{\rm
 cav}\rho,
 \ee
where now
 \bma
 \label{Hab}
 \bea
 \tilde H_a &=& g (\tau_a^+ \tilde a + \tilde a^\dagger
 \tau_a^-),\\
 \tilde H_b &=& g (\tau_b^+ \tilde b + \tilde b^\dagger
 \tau_b^-),\nonumber\\
 \tilde {\cal L}_{\rm cav}\rho&=&\kappa \sum_{\alpha=\tilde a,\tilde b} (
 \alpha\rho\alpha^\dagger- \alpha^\dagger\alpha\rho),
 \eea
 \ema
with $\tau_{a,b}^+=(\tau_{a,b}^-)^\dagger=(N+1)^{1/2}
\sigma_{a,b}^+ - N^{1/2} \sigma_{b,a}^-$. Solving master equation
(\ref{ME2}) seems to be a difficult task. However, one can easily
determine the steady state, which is given by
 \be
 \label{ss}
 |\Psi\rangle = \left(\sqrt{\frac{N+1}{2N+1}}
 |g\rangle_a|g\rangle_b +
 \sqrt{\frac{N}{2N+1}}|e\rangle_a|e\rangle_b\right)
 |0\rangle_{\tilde a}|0\rangle_{\tilde b},
 \ee
where $|0\rangle_{\tilde a,\tilde b}$ are the vacuum states of the
new modes $\tilde a$ and $\tilde b$, respectively. This is a pure
state, which in the limit $N\gg 1$ tends to a maximally entangled
state. For a realistic value of $N\sim 1$ one still obtains a
state with a large entanglement of formation (EoF) $E(\Psi)\sim
0.92$.

After the creation of the state (\ref{ss}), one should
simultaneously switch off the squeezing source and transfer the
excited state $|e\rangle$ of both atoms to some other internal
ground state $|g'\rangle$ using a laser, in order to avoid
spontaneous emission [Fig.\ \ref{fig2}(a), (i)]. Once this is
done, a maximally entangled state can be created as follows [Fig.\
\ref{fig2}(a), (ii)]. In each of the atoms, a radio frequency (or
two--photon Raman) pulse is applied which transforms
$|g'\rangle\to cos\theta |g'\rangle + \sin\theta |g''\rangle$,
where $|g''\rangle$ is an auxiliary internal ground state, while
the state $|g\rangle$ is not affected by the pulse. Then, the
state $|g''\rangle$ is detected in both atoms using the quantum
jump technique. If the outcome is negative, one can easily show
that the atomic state will be projected onto a state proportional
to $|g\rangle_a|g\rangle_b + |g'\rangle_a|g'\rangle_b$ if one
chooses $\cos(\theta)=[(N+1)/N]^{1/4}$. Note that this measurement
corresponds to a generalized measurement but in which the role of
the ancilla is taken by the auxiliary level $|g''\rangle$, i.e. no
extra atoms are required. The success probability depends on the
value of $N$, but after a sufficiently large number of trials, a
maximally entangled state can be prepared for any value of $N>0$.

In practice, there will be several physical phenomena which will
distort the atomic entanglement in steady state. In the following,
we will evaluate the effect of the most important sources of
imperfection.

In order to analyze the non--ideal situation in which $\Gamma\ne
0$ and $M< [N(N+1)]^{1/2}$, we consider the limit
$g\sqrt{N+1},\Gamma \ll \kappa$. Then, we can eliminate the cavity
mode by generalizing the procedure presented in \cite{Cirac2}. We
define $\sigma:={\rm tr}_{a,b}(\rho)$, so that
 \be
 \label{ME3b}
 \frac{d\sigma}{dt}= {\cal L}_1 \rho +
 {\cal L}_{\rm at}^{a} \sigma + {\cal L}_{\rm at}^{b}\sigma,
 \ee
where ${\cal L}_1(\rho)=-ig {\rm
tr}_{a,b}(a[\sigma_a^+,\rho]-h.c.) + a\leftrightarrow b$. On the
other hand, integrating formally Eq.\ (\ref{ME}), and substituting
the result in (\ref{ME3b}) one can check that in the limit $\kappa
t\gg 1$, the dominant contribution is given by the term coming
from
 \be
 \rho(t)\simeq \int_0^t d\tau e^{{\cal L}_{\rm cav} \tau}
 {\cal L}_2 [\rho(t-\tau)],
 \ee
where ${\cal L}_2(\rho)=-ig ([a,\rho\sigma_a^+]-h.c.) +
a\leftrightarrow b$. Using that $e^{{\cal L}_{\rm cav}
\tau}([a,R])=e^{-\kappa \tau} [a,e^{{\cal L}_{\rm cav} \tau}R]$ we
see that the integrand will vanish for times $\kappa \tau\gg 1$,
so that we can extend the limit of the integral to infinity.
Moreover, since after the time $t$ the cavity mode will be driven
to its steady state, $\rho_{ss}$, which fulfills ${\cal L}_{\rm
cav}(\rho_{ss})=0$, we can replace $e^{{\cal L}_{\rm cav}
\tau}\rho(t-\tau)\to \sigma(t)\otimes \rho_{ss}$. This procedure
amounts to performing the standard Born--Markov approximations
\cite{GardinerZoller}, but here we have that the bath itself
(cavity mode) undergoes a dissipative dynamics. After some lengthy
algebra we obtain
 \bea
 \label{ME3}
 \dot\sigma&=&
 \frac{\gamma}{2} (n+1) \sum_{\alpha=a,b} (
 \sigma_\alpha^-\sigma\sigma_\alpha^+- \sigma_\alpha^+\sigma_\alpha^-\sigma) \nonumber\\
 &&+ \frac{\gamma}{2} n \sum_{\alpha=a,b} (
 \sigma_\alpha^+\sigma\sigma_\alpha^-- \sigma_\alpha^-\sigma_\alpha^+\sigma) \nonumber\\
 &&+ \gamma m (\sigma_a^-\sigma \sigma_b^- + \sigma_b^-\sigma \sigma_a^- -
 \sigma_b^-\sigma_a^-\sigma -\sigma_a^-\sigma_b^-\sigma)
 \nonumber\\
 &&+ h.c.
 \eea
Here
 \bma
 \bea
 \label{nm}
 \gamma&=& \frac{g^2}{\kappa}(2+\epsilon),\quad\epsilon:=\Gamma\kappa/(g^2) \\
 n&=& N (1+\epsilon/2)^{-1},\quad m=-M (1+\epsilon/2)^{-1}.
 \eea
 \ema

The interpretation of master equation (\ref{ME3}) is
straightforward. It describes the interaction of the two atoms
with a common squeezed reservoir in which the squeezing parameters
are renormalized due to the presence of spontaneous emission. The
steady state solution only depends on $n$ and $m$, and can be
easily determined. In fact, for $\Gamma=0$ and perfect squeezing
we recover the steady state (\ref{ss}), as expected. Instead of
analyzing our results in terms of $n$ and $m$, it is more
convenient to analyze them in terms of the physical parameters
$\epsilon$ and $N$, choosing (\ref{perfectsq}). Note that it is
always possible to find an $\epsilon$, and an $N$ and $M$
fulfilling (\ref{perfectsq}), which give any prescribed values of
$n$ and $m$, so that the effects of imperfect squeezing can be
directly read off from our analysis.

\begin{figure}[t]
  \centering
  \resizebox{\linewidth}{!}{
  \includegraphics{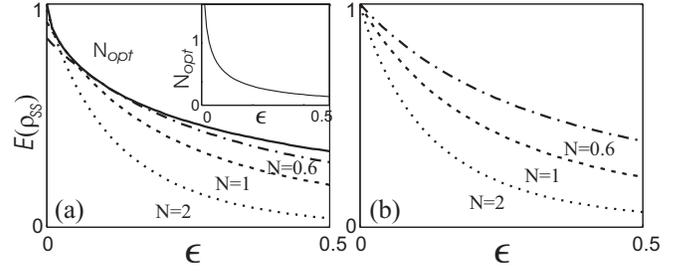}}
  \caption{\label{fig3}
    EoF of the atoms in steady state
    as a function of the parameter $\epsilon$ and various values of $N$:
    (a) Without generalized measurement; (b) With generalized
    measurement. The solid line indicates the optimal value and the insert
    gives the $N$ for which the EoF is optimal.}
\end{figure}

In Fig.\ [\ref{fig3}(a)] we have plotted the atomic EoF of the
steady state as a function of $\epsilon$ for various values of
$N$. The most important aspect is that for $\epsilon\ne 0$
increasing the squeezing does not necessarily lead to an increase
in the EoF. For each value of $\epsilon$ we have determined the
best choice of $N$, which is shown in the insert. For realistic
parameters $\epsilon\alt 0.1$ the best choice of $N$ is around
$0.6$, leading to an EoF of $0.638$. In Fig.\ [\ref{fig3}(b)] we
have plotted the results when the filtering measurement described
above is performed. Here we see that the achievable entanglement
significantly increases. For example, for $\epsilon=0.1$ one
obtains $0.775$.

\begin{figure}[t]
  \centering
  \resizebox{\linewidth}{!}{
  \includegraphics{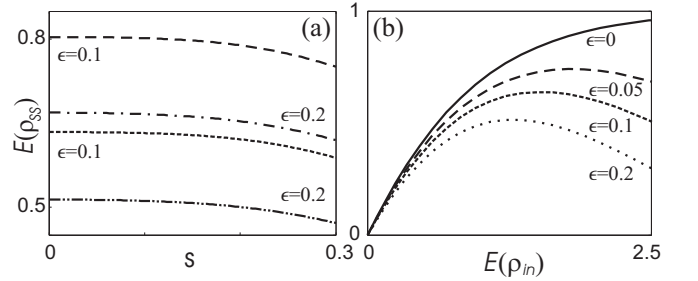}}
  \caption{\label{fig4}
    (a) EoF of the atoms in steady state
    as a function of the parameter $s$, for $N=0.5$.
    Upper (lower) two curves correspond to
    the case with (without) measurement after preparation;
    (b) EoF of the atomic state as a function of the one
    corresponding to the squeezed state.}
    \end{figure}

In Fig.\ [\ref{fig4}(a)] we have analyzed the effects of the
imprecision in the position of the atoms \cite{noteLDL}. To this
aim, we have first extended our analysis to the case
$g_a=g\cos(\theta_a)\ne g_b=g\cos(\theta_b)$, by deriving a master
equation analogous to (\ref{ME3}). We have then averaged the
density operator corresponding to the steady state with respect to
$\theta_a$ and $\theta_b$, with a weight function
$p(\theta)\propto \exp[-\theta^2/(2s^2)]$. We have plotted the
resulting EoF as a function of $s$, which measures the
experimental uncertainty in the position of the ion. The figure
shows that this uncertainty does not have dramatic effects in the
EoF, as long as the position of the particle is not far from the
antinode of the cavity mode standing wave.

As mentioned in the introduction, with this scheme we are
transforming the continuous variable entanglement contained in the
squeezed vacuum state of the incident light into discrete (qubit)
entanglement. In Fig.\ [(\ref{fig4}(b)] we have analyzed the
efficiency of this process. We have plotted the achieved EoF as a
function of the EoF contained in the squeezed state for various
values of $\epsilon$. The transfer is more efficient for low
values of $N$, something that can be attributed to the fact that
only two Schmidt coefficients are relevant for the two--mode
squeezed state.

An important aspect of our scheme is that it can be scaled up to
build a quantum communication network or quantum repeaters. The
idea is to embed two (or more) atoms in each cavity, and to use
two modes in each of them. Atoms A1 and B2 can interact with modes
$a_1$ and $b_2$ in their respective cavities, which in turn are
driven by two--mode squeezed light. Atoms B1 and C2 can also
become entangled in a similar way by interacting with modes $b_1$
and $c_2$, respectively. In the ideal case, after the entanglement
is created, a measurement in atoms B1 and B2 will yield an
entangled state between atoms A1 and C2. In the presence of
imperfections, the entanglement will be degraded every time we
perform one of these operations (i.e. as we try to extend the
entanglement over longer distances). In order to avoid this
problem, one can use other auxiliary atoms in each cavity and
perform entanglement purification as it is required to build a
quantum repeater \cite{repeaters}.

In the case of a small number of modes, it is possible to perform
these experiments with a single atom per cavity and without having
to perform joint measurements. This is not possible with
two--level atoms, since it is known that in that case there is a
maximum amount of entanglement that it can share with two
neighboring atoms \cite{Duer}.  This problem can be circumvented
by using several internal states, since in that case it is indeed
possible that one atom shares two ebits with two other atoms. For
example, one may take the atomic scheme of Fig.\ \ref{fig2}(b). We
have renamed the internal state since then it is simpler to
understand the scheme. Two cavity modes are used, that connect
pairs of levels with the help of off--resonant laser beams in
Raman configuration. Now, let us consider that we have three atoms
A, B, C, in three different cavities. The atoms in A and C have
the same configuration as before, whereas the atom in cavity B has
the one indicated in Fig.\ \ref{fig2}(b). The Hamiltonian, after
adiabatically eliminating the excited state of atom B has the form
 \be
 H= g(\sigma_a^+ a + \sigma_{b_1}^+ b_1 +
 \sigma_{b_2}^+ b_2 + \sigma_c^+ c) + h.c.
 \ee
Here, $\sigma_{a,c}^+$ are defined as before, whereas
 \bma
 \bea
 \sigma_{b_1}^+ &=& |1,0\rangle_B\langle 0,0| + |1,1\rangle_B\langle
 0,1|,\\
 \sigma_{b_2}^+ &=& |0,1\rangle_B\langle 0,0| + |1,1\rangle_B\langle
 1,0|.
 \eea
 \ema
Now, if modes $a$ and $b_1$ and modes $c$ and $b_2$ are driven by
two independent sources of squeezed light, it is easy to check
that under ideal conditions ($\Gamma=0$ and perfect squeezing) the
atomic steady state is
 \bea
 |\Psi\rangle_{ss} &=& \frac{N+1}{2N+1}|g\rangle_A|0,0\rangle_B|g\rangle_C +
 \frac{N}{2N+1}|e\rangle_A|1,1\rangle_B|e\rangle_C\nonumber\\
 &+& \frac{\sqrt{N(N+1)}}{2N+1}(|g\rangle_A|0,1\rangle_B|e\rangle_C
 + |e\rangle_A|1,0\rangle_B|g\rangle_C)\nonumber.
 \eea
In the limit $N\gg 1$ this state contains two ebits, one between A
and B and another between B and C. Alternatively, an appropriate
measurement in B will produce a maximally entangled state between
A and C with certain probability. This scheme can be easily
generalized to a larger number of nodes. However, as mentioned
above, the role of the imperfections will be important and one
eventually needs to consider several atoms in each cavity to
purify the obtained entanglement.

In conclusion, we have shown that atoms can get entangled by
interacting with a common source of squeezed light. The continuous
variable entanglement can, in this way, be transformed in discrete
one in steady state. Local measurements result in a more efficient
entanglement creation. Given the experimental progress in trapping
atoms inside cavities \cite{Kimble3} and the successful
experiments on coupling squeezed light into a cavity
\cite{Kimble}, the present scheme may become a very robust
alternative to current methods to construct quantum networks for
quantum communication.

This work has been supported in part by the
\emph{Kompetenznetzwerk Quanteninformationsverarbeitung der
Bayerischen Staatsregierung} and by the EU IST projects "RESQ" and
"QUPRODIS". After completion of this work we have learned of a
related problem but using an atomic Raman
configuration\cite{Parkins3}.


\begin{thebibliography}{99}
\bibitem{noteLDL} We are assuming that the atoms are confined in
the LDL, so that their motion does not play any role. It may be
interesting to detune the cavity mode to the red to enhance
sideband cooling.
\end{thebibliography}
\end{document}